\documentclass[fleqn,11pt]{article}

\usepackage{amssymb,amsmath}

\usepackage{amsfonts,amssymb,cite}
\usepackage{graphicx}

\def\noi{\noindent}

\newcommand{\Title}[1]{\noi {{\Large\bf #1}}\\[1ex]}

\def\Aunames#1{\noi{\bf #1}}
\def\au#1{${}^{#1}$}
\def\Addresses#1{\medskip\noi \protect
	\begin{description}\itemsep -3pt {\it #1} \end{description}}
\def\adr#1#2{\item[${}^{#1}$]{\it #2}}

\newcommand{\Abstract}[1]{\vskip 2mm \begin{center}
        \parbox{16.4cm}{\small\noi #1} \end{center}\medskip}

\def\email#1#2{\footnotetext[#1]{e-mail: #2}\addtocounter{footnote}{1}}

\def\nq{\hspace*{-1em}}
\def\nqq{\hspace*{-2em}}
\def\nhq{\hspace*{-0.5em}}

\def\qq{\qquad}
\def\cm{\hspace*{1cm}}
\def\inch{\hspace*{1in}}

\usepackage{color}

\def\Funding#1{\subsection*{Funding} #1}

\def\ConflictThey{\subsection*{Conflict of interest} 
	The authors declare that they have no conflicts of interest.}

\def\Jl#1#2{#1 {\bf #2},\ }

\def\ApJ#1 {\Jl{Astroph. J.}{#1}}
\def\CQG#1 {\Jl{Class. Quantum Grav.}{#1}}
\def\DAN#1 {\Jl{Dokl. AN SSSR}{#1}}
\def\GC#1 {\Jl{Grav. Cosmol.}{#1}}
\def\GRG#1 {\Jl{Gen. Rel. Grav.}{#1}}
\def\IJMPD#1 {\Jl{Int. J. Mod. Phys. D}{#1}}
\def\JETF#1 {\Jl{Zh. Eksp. Teor. Fiz.}{#1}}
\def\JETP#1 {\Jl{Sov. Phys. JETP}{#1}}
\def\JHEP#1 {\Jl{JHEP}{#1}}
\def\JMP#1 {\Jl{J. Math. Phys.}{#1}}
\def\NPB#1 {\Jl{Nucl. Phys. B}{#1}}
\def\NP#1 {\Jl{Nucl. Phys.}{#1}}
\def\PLA#1 {\Jl{Phys. Lett. A}{#1}}
\def\PLB#1 {\Jl{Phys. Lett. B}{#1}}
\def\PRD#1 {\Jl{Phys. Rev. D}{#1}}
\def\PRL#1 {\Jl{Phys. Rev. Lett.}{#1}}

\def\lal{&&\nqq {}}
\def\eq{Eq.\,}
\def\eqs{Eqs.\,}
\def\beq{\begin{equation}}
\def\eeq{\end{equation}}
\def\bear{\begin{eqnarray}}
\def\bearr{\begin{eqnarray} \lal}
\def\ear{\end{eqnarray}}
\def\earn{\nonumber \end{eqnarray}}

\def\nnn{\nonumber\\ \lal }
\def\nnnv{\nonumber\\[5pt] \lal }
\def\yy{\\[5pt] {}}
\def\yyy{\\[5pt] \lal }

\def\D{\partial}

\def\const{{\rm const}}
\def\eps{\varepsilon}

 \catcode`\@=11 \@addtoreset{equation}{section}\catcode`\@=12

\begin{document}
\twocolumn[
%\jnumber{3}{2024}

\Title{On a reconstruction procedure for special spherically\yy symmetric metrics
	 in the scalar-Einstein-Gauss-Bonnet model:\yy the Schwarzschild metric test  }

\Aunames{K. K. Ernazarov\au{a} and V. D. Ivashchuk\au{a,b,1} } 

\Addresses{\small
\adr a {Peoples' Friendship University of Russia (RUDN University), 
             ul. Miklukho-Maklaya 6, Moscow 117198, Russia}         
\adr b {Center for Gravitation and Fundamental Metrology, VNIIMS,
            ul. Ozyornaya  46, Moscow 119361, Russia} 
	}

%\Dates{May 22, 2024}{May 24, 2024}{May 28, 2024}

\Abstract
	{ The 4D gravitational model with a real scalar field $\varphi$, Einstein and  
	Gauss-Bonnet terms is considered. The action contains the potential $U(\varphi)$ and 
	the Gauss-Bonnet coupling function $f(\varphi)$.	For a special static spherically 
	symmetric metric  $ds^2 = (A(u))^{-1} du^2 - A(u) dt^2 + u^2 d\Omega^2$,
	with $A(u) > 0$  ($u > 0$ is a radial coordinate), we verify the so-called reconstruction
	procedure suggested by Nojiri and Nashed. This procedure presents certain implicit
	relations for $U(\varphi)$ and $f(\varphi)$ which lead  to exact solutions 
      to the equations of motion for a given  metric governed by $A(u)$. We confirm that 
      all relations in the approach of  Nojiri and  Nashed for $f(\varphi(u))$ and $\varphi(u)$ 
      are correct, but the relation for $U(\varphi(u))$ contains  a typo which is eliminated 
      in this paper.  Here we apply the procedure to the (external) Schwarzschild metric with
      the gravitational radius $2 \mu $ and $u > 2 \mu$. Using the  ``no-ghost'' restriction 
      (i.e., reality of $\varphi(u)$), we find two families of $(U(\varphi), f(\varphi))$. The first 
      one gives us the Schwarzschild metric defined for $u > 3 \mu$, while the second one
      describes the Schwarzschild metric defined for $2 \mu < u < 3 \mu$ ($3 \mu$ is the 
      radius of the photon sphere). In both cases the potential $U(\varphi)$ is negative. }
    
\medskip    
]    %%%%%%%%%%%%%%%%%%%%%%%%%%%%%%%%%%%%%%%%%%%
\email 1 {ivas@vniims.ru}

{ % au-def
\def\R{{\mathbb R}} 

% ==================
\section{Introduction}
% ==================

  The pursuit of a unified description of gravity with quantum mechanics has driven
  theoretical physics for decades. String theory, which was conjectured to be a promising 
  candidate for this unification, ''predicted'' the existence of higher-dimensional space-time 
  and a plethora of new fields, including the scalar dilaton. String theory also predicted, in 
  the low energy limit, certain extensions of General Relativity (GR).  One such extension 
  involves incorporating the Gauss-Bonnet (GB) term \cite{iv-Zwiebach,iv-FrTs1,iv-FrTs2,iv-GrW}, 
  coupled to a function of a scalar field (dilaton), leads to a rich and complex landscape of 
  scalar-Einstein-Gauss-Bonnet (sEGB) gravity. We note that the pure GB term gives us 
  a topological invariant in four dimensions while it is dynamically relevant in higher dimensions. 

  The advent of sEGB gravity challenges the conventional understanding of black holes
  established by GR. A nontrivial coupling between the scalar field and the GB term leads 
  to deviations from the Schwarzschild solution, ushering in a new ``era'' of ``hairy'' black 
  holes characterized by scalar hair. This scalarization, studied extensively by Kanti et al. 
  \cite{iv-1,iv-2} and other authors, has profound implications for the properties of black 
  holes, influencing their stability, computability, thermodynamics, and interaction with
  the surrounding matter --- see \cite{iv-3Kon,iv-3Klei} and references therein. J. Kunz et al. 
  and some other authors extensively studied static and rotating black hole solutions 
  in this model, revealing their unique characteristics \cite{iv-3Klei}. These black holes 
  possess a scalar charge, which affects their gravitational field and thermodynamic properties.
  The paper by Bronnikov and Elizalde \cite{iv-BronEl} made an important 
  contribution to the theoretical description of possible black hole configurations in the 
  sEGB model (with a scalar field potential term): it was found that the GB term, in general, 
  violates certain well-known ``no-go'' theorems, which are valid for a minimally coupled 
  scalar field in GR.  

  While the theoretical foundations of sEGB gravity are compelling, observational evidence
  remains crucial for validating its predictions. Fortunately, sEGB black holes exhibit distinct
  observational signatures that can be detected through various astrophysical probes. One 
  such probe involves gravitational waves. Merging black holes in sEGB gravity are expected 
  to emit gravitational waves with characteristic deviations from GR predictions. The possible 
  detection and analysis of these gravitational waves by detectors like LIGO and Virgo offer
  a powerful tool for testing the validity of sEGB gravity and constraining the parameters 
  of the model \cite{iv-GW}.

  Another promising avenue for probing sEGB black holes lies in studying their shadows 
  \cite{iv-PerTs} and quasinormal modes \cite{iv-KonZh}. The shadow of a black hole, a dark
  silhouette against a bright background, is influenced by the black hole's geometry and the
  surrounding space-time. As shown by Cunha et al. [7], sEGB black holes exhibit distinctive
  shadow morphologies, deviating from the circular shadows predicted by GR. Similarly, the
  quasinormal modes of black holes, characteristic frequencies emitted during perturbations, 
  are also sensitive to the presence of the scalar field and the GB term \cite{iv-8}. These
  observational signatures offer unique opportunities to distinguish sEGB black holes from 
  their GR counterparts.  
  
  This paper is inspired by the recent article of Nojiri and  Nashed \cite{iv-NN}, which 
  delves into the realm of special spherically symmetric black holes within the sEGB 
  model governed by the coupling function $f(\varphi)$ and the potential function 
  $V(\varphi)$, where $\varphi$ is a scalar field.
   In Ref.\,\cite{iv-NN}, the authors were dealing with special static spherically symmetric 
   metric
\[
	ds^2 = \left(a(r)\right)^{-1}dr^2 - a(r)dt^2 + r^2 d\Omega^2. 
\]	
  They have solved (partly) the reconstruction problem: for a given redshift function 
  $a(r) > 0$, they  found implicit relations for $f(\varphi)$ and $V(\varphi)$, which lead  
  to exact solutions  to the equations of motion with the given metric. The problem was 
  solved up to (global) resolution of the ghost avoiding restriction, coming from the reality
  condition for the scalar field solution $\varphi(r)$. Here we verify all reconstruction 
  relations from \cite{iv-NN}, and after eliminating a typo in the relation for  $V(\varphi)$ 
  we apply the reconstruction procedure to the simplest case of the Schwarzschild metric. 
  In this case, the ghost avoiding problem may be readily solved.  

% =====================================
\section{The  scalar-Einstein-Gauss-\\Bonnet model}
% =====================================

  We are dealing with the so-called scalar-Einstein-Gauss-Bonnet model which is governed 
  by the action
\bearr
		S = \int d^4 z\   \sqrt{|g|} \bigg(\frac{R\big(g\big)}{2\kappa^2} 
			- \frac{1}{2}g^{MN}\D_M\varphi\D_N\varphi   \label{iv-1A} 
\nnn \inch  
		 - U(\varphi)  + f(\varphi) {\cal G}  \bigg),
\ear
  vvwhere  $\kappa^2 = 8\pi {G}/{c^4}$, $\varphi$ is a scalar field, 
  $g_{MN}dz^M \otimes dz^N$ is the 4D metric, 
  $R[g]$ is the scalar curvature, $\cal{G}$ is the Gauss-Bonnet invariant, 
  $U(\varphi)$ is potential, and $f(\varphi)$ is a coupling function.

  We study spherically-symmetric solutions with the metric
\bearr
		ds^2 = g_{MN}dz^Mdz^N
\nnn \qq 
	  = e^{2\gamma(u)}du^2 - e^{2\alpha(u)}dt^2 + e^{2\beta(u)}d\Omega^2,  \label{iv-2A} 
\ear
  defined on the manifold
\beq  
   		M = \R \times \R_{*} \times S^2.    \label{iv-3A}
\eeq  
  Here $\R_{*} = \left(2\mu, +\infty \right)$, and $S^2$ is a 2D sphere with the metric 
  $d\Omega^2 = d\theta^2 + \sin^2\theta d\varphi^2$, where $ 0 < \theta < \pi$, 
  and $0 < \varphi < 2\pi$.

  By substituting the metric  (\ref{iv-2A}) into the action we obtain 
  $S = 4\pi \int du  (L + {d ( \dots )}/{du} )$,  where the Lagrangian $L$ reads
\bearr
	L = \frac{1}{\kappa^2}\bigg[e^{\alpha- \gamma + 2\beta}\dot{\beta}
		\Big(\dot{\beta}+ 2\dot{\alpha}\Big) + e^{\alpha + \gamma}\bigg]
\nnn \cm
	  - \frac{1}{2}e^{\alpha - \gamma + 2\beta}\dot{\varphi}^2
			     - e^{\alpha + \gamma + 2\beta}U(\varphi)
\nnn \cm
      - 8\dot\alpha\dot\varphi\frac{df}{d\varphi}
      \Big(\dot{\beta}^2e^{\alpha + 2\beta  - 3\gamma}
        - e^{\alpha - \gamma}\Big), 				\label{iv-4A} 
\ear
  and the total derivative term ${d ( \dots ) }/{du}$ is irrelevant for our consideration.

  Here and in what follows we denote $\dot{x} = {dx}/{d u}$. 
  The equations of motion for the action (\ref{iv-1A})  with the metric (\ref{iv-2A}) 
  involved are equivalent to the Lagrange equation corresponding to the Lagrangian (\ref{iv-4A}).

  The Lagrange equations read 
\bearr
	\frac{\D L}{\D \gamma} = \frac{1}{\kappa^2}
	\bigg[- e^{\alpha - \gamma + 2\beta}\dot{\beta}
		\left(\dot{\beta} + 2\dot{\alpha}\right) + e^{\alpha + \gamma}\bigg]
\nnn \cm 
	 + \frac{1}{2}e^{\alpha - \gamma + 2\beta}\dot{\varphi}^2 
 			-  e^{\alpha + \gamma + 2\beta}U(\varphi) 
\nnn  \qq
	-  8\dot{\alpha}\dot{\varphi}\frac{df}{d\varphi}\bigg(-3 \dot{\beta}^2e^{\alpha 
	+ 2\beta - 3\gamma} + e^{\alpha - \gamma}\bigg) \nonumber \\
\nnn  \qq	 = 0, 		\label{iv-5G}
\ear
\bearr
	\frac{d}{du}\bigg(\frac{\D L}{\D \dot{\alpha}}\bigg)  - \frac{\D L}{\D \alpha}  
\nnn \nq
	 =\! \frac{d}{du}\!\bigg[\frac{2}{\kappa^2} e^{\alpha - \gamma + 2\beta}\dot{\beta} 
	 - 8\dot{\varphi}\frac{df}{d\varphi} \Big(\dot{\beta}^2
	 e^{\alpha + 2\beta - 3\gamma} -\!  e^{\alpha - \gamma}\Big) \bigg] 
\nnn 
	 - \bigg[ \frac{1}{\kappa^2} \left(e^{\alpha - \gamma + 2\beta}\dot{\beta}
	 \big(\dot{\beta} + 2\dot{\alpha}\big) + e^{\alpha + \gamma} \right) 
\nnn \cm
 	- \frac{1}{2}e^{\alpha - \gamma + 2\beta}\dot{\varphi}^2 
 	- e^{\alpha + \gamma + 2\beta}U(\varphi) 
\nnn \cm
	 - 8\dot{\alpha}\dot{\varphi}\frac{df}{d\varphi}\Big(\dot{\beta}^2
		 e^{\alpha + 2\beta - 3\gamma} - e^{\alpha - \gamma}\Big) \bigg] = 0,      \label{iv-5A}  
\ear
\bearr
		\frac{d}{du}\bigg(\frac{\D L}{\D \dot{\beta}}\bigg)  - \frac{\D L}{\D \beta}  
\nnn 
 	= \frac{d}{du}\bigg[\frac{1}{\kappa^2} e^{\alpha - \gamma
		  + 2\beta}\big(2\dot{\beta} + 2\dot{\alpha}\big) 
\nnn  
	  - 8\dot{\alpha}\dot{\varphi}\frac{df}{d\varphi}2\dot{\beta}
  			e^{\alpha + 2\beta - 3\gamma}\bigg] 
  			- \bigg[\frac{2}{\kappa^2}e^{\alpha - \gamma + 2\beta}\dot{\beta}
 			\big(\dot{\beta} + 2\dot{\alpha}\big) 
\nnn \cm
 	-  e^{\alpha - \gamma + 2\beta}\dot{\varphi}^2
   	- 2e^{\alpha + \gamma + 2\beta}U(\varphi) 
\nnn   	\cm
		 - 16\dot{\alpha}\dot{\varphi}\frac{df}{d\varphi}\dot{\beta}^2
		 e^{\alpha + 2\beta - 3\gamma} \bigg]  = 0,      			\label{iv-5B} 
\ear
  and
\bearr
		\frac{d}{du}\Bigg(\frac{\D L}{\D \dot{\varphi}}\Bigg) - \frac{\D L}{\D \varphi}
		= \frac{d}{du}\bigg[ - e^{\alpha - \gamma + 2\beta}\dot{\varphi}
\nnn\cm
	  - 8\dot{\alpha}\frac{df}{d\varphi}\left(\dot{\beta}^2e^{\alpha + 2\beta - 3\gamma}
			   - e^{\alpha - \gamma}\right)\bigg]
\nnn
	 + \bigg[e^{\alpha + \gamma + 2\beta}\frac{dU}{d\varphi}
\nnn	 \ \ \ 
    +  8\dot{\alpha}\dot{\varphi}\frac{d^2f}{d\varphi^2}\left(\dot{\beta}^2
    e^{\alpha + 2\beta - 3\gamma} - e^{\alpha - \gamma}\right)\bigg] =0.         \label{iv-5P}
\ear

% =============================
\section{The reconstruction procedure}
% =============================
 
  As in Ref. \cite{iv-NN}, we consider a special ansatz for the metric (\ref{iv-2A}),
\beq  
		ds^2 =  \frac {du^2}{A(u)} - A(u)dt^2 + u^2 d\Omega^2,    \label{iv-5Buch} 
\eeq  
  where
\bearr
	e^{2\gamma(u)} = 1/A(u), \qq   e^{2\alpha(u)} = A(u) > 0,
\nnn \cm  
		e^{2\beta(u)} = u^2 > 0.                       \label{iv-5AC}
\ear
   In what follows we use the identities
\bear
		\dot{\alpha}= \frac{\dot{A}}{2A}, \qquad  \dot{\beta}= \frac{1}{u}.  \label{iv-5AB}
\ear
  As was done in Ref. \cite{iv-NN}, we put without loss of generality $\kappa^2 = 1$. 
  We also denote
\beq  
 	  f \left(\varphi(u)\right) = f\left(u \right), 
		  \qquad  U \left(\varphi(u)\right) = U\left(u \right),  \label{iv-6_FU}
\eeq  
  and hence,
\bearr
	\frac{d}{du}f(u) = \frac{df}{d\varphi}\frac{d\varphi}{du} 
	\ \Longleftrightarrow \ \dot{f} = \frac{df}{d\varphi}\dot{\varphi},  \label{iv-6_F} 
\yyy
	\frac{d}{du}U(u) = \frac{dU}{d\varphi}\frac{d\varphi}{du} 
	\ \Longleftrightarrow\  \dot{U} = \frac{dU}{d\varphi}\dot{\varphi}.  \label{iv-6_U} 
\ear
  Strictly speaking, one should use other notations in (\ref{iv-6_FU}), for instance: 
  $f(\varphi(u)) = \hat{f}(u)$, $U(\varphi(u)) = \hat{U}(u)$. We hope that notations 
  in (\ref{iv-6_FU}) will not lead to a confusion.

  Multiplying (\ref{iv-5G}) by $(-2)$ and using the relations (\ref{iv-5AB}), (\ref{iv-6_F}), we get
\bearr  
		\dot{A}\left[8\dot{f}\left(1 - 3A\right) + 2 u \right]  + 2A - 2
\nnn \inch		
		 - u^2 A\dot{\varphi}^2 + 2 u^2 U = 0.  \label{iv-6G} 
\ear  
  Equation (\ref{iv-6G}) coincides with \eq (10) from Ref.\,\cite{iv-NN}.

  Multiplying (\ref{iv-5G}) by (-2) and using the relations (\ref{iv-5AB}) and (\ref{iv-6_F}), we get
\bearr
		16\ddot{f}A\left(1 - A\right) + 8\dot{f} \left(\dot{A} - 3A\dot{A} \right) +2u \dot{A} 
\nnn \cm  
		 + 2A + u^2 A\dot{\varphi}^2 - 2 + 2 u^2 U = 0.  \label{iv-6A} 
\ear
    Equation (\ref{iv-6A}) coincides with \eq (9) from \cite{iv-NN}.

   Analogously, using (\ref{iv-5AB}) and (\ref{iv-6_F}), we rewrite \eq (\ref{iv-5B}) as 
\bearr
	\left(u^2 - 8u \dot{f}A \right)\ddot{A} - 8u \ddot{f}\dot{A}A
\nnn \cm 	
 - 4\dot{f} \left(\dot{A}^2 u^2 + 2 \dot{A}A - 2\dot{A}A \right) 
\nnn \cm 
    +  2u\dot{A} + u^2 \left(A\dot{\varphi}^2 + 2U\right)  = 0. \qquad \label{iv-6B} 
\ear
   Equation (\ref{iv-6B}) coincides with \eq (11) from \cite{iv-NN}.

  Now, multiplying \eq (\ref{iv-5P}) by $(-\dot{\varphi})$, we obtain
\bearr
		4\dot{f}\left(A - 1\right)\ddot{A} + \ddot{\varphi}
			\dot{\varphi}A u^2 + 4\dot{f}\dot{A} \dot{A} 
\nnn \cm 
		+ \left(\dot{A} u^2 + 2u A \right)\dot{\varphi}^2 - u^2 U = 0.  \label{iv-6P} 
\ear

  In the case where
\bear
		\dot{\varphi} \neq 0 \quad {\rm for} \quad u \in \left(u_{-}, u_{+}\right), \label{iv-7PN} 
\ear
  in some interval $\left(u_{-}, u_{+}\right)$ belonging to $\mathbb{R}$, the relations 
  (\ref{iv-6P})  and (\ref{iv-5P}) are equivalent in this interval. Equation (\ref{iv-6P})
  coincides with \eq (12) from Ref. \cite{iv-NN}. 

  By adding \eqs (\ref{iv-6A}) and (\ref{iv-6G}) and dividing the result by 4, 
  we get the expression for the potential function $U = U(u)$
\bearr
		U = \frac{1}{u^2}\bigg[1 - 4A\left(1 - A\right)\ddot{f} 
\nnn \cm  
	- \dot{A}\left[4\dot{f}\left(1 - 3A\right) +  u \right]  - A  \bigg].  \label{iv-6U} 
\ear
  Here we note that \eq (\ref{iv-6U}) coincides with \eq (13) from \cite{iv-NN} up to a typo: 
  in \eq (13) from \cite{iv-NN} the term $a'$ in square brackets should be omitted. 

  The relation (\ref{iv-6U}) may be written as
\bear
		 u^2 U = E_{U}\ddot{f} + F_{U}\dot{f} + G_U , \label{iv-6UU} 
\ear
  where
\bearr
		E_U = -4A \left(1 -  A\right),  \label{iv-6EU} 
\yyy
		F_U =  -4\dot{A}\left(1 - 3  A\right), \label{iv-6FU} 
\yyy		
		G_U = 1 - \dot{A} u -  A. \label{iv-6GU} 
\ear
  Subtracting (\ref{iv-5G}) from (\ref{iv-5A}) and dividing the result by $2A$, 
  we obtain a relation for $\dot{\varphi}$:
\bear
	 \dot{\varphi}^2 = 8\ddot{f}\left(A - 1 \right) u^{-2}  \equiv \Phi. \label{iv-6P2} 
\ear
  This relation coincides with \eq (14) from \cite{iv-NN}. 
  Due to  (\ref{iv-7PN}) and $u > 0$, we get  a ghost avoiding restriction (GAC)
  explored in \cite{iv-NN}, 
\bear
		\Phi = \Phi(u) > 0 \label{iv-6Phi} 
\ear
  for all $u \in \left(u_{-}, u_{+}\right)$.
 
  Subtracting (\ref{iv-5A}) from (\ref{iv-5B}), we get the master equation for the 
  coupling function $f = f(u)$:
\beq  
		  E\ddot{f} + F\dot{f} + G = 0,  \label{iv-6FF} 
\eeq  
  where 
\bearr
	 E =  8A\left(2A - u \dot{A} - 2\right), 		\label{iv-6E} 
\nnn	 
		F = -8u\ddot{A}A  - 8u \dot{A}^2  + 8\left(3A 
			 - 1 \right)\dot{A}, 		 \label{iv-6F6} 
\nnn
		G = u^2 \ddot{A} - 2A + 2.			\label{iv-6G6} 
\ear
  The master equation (\ref{iv-6FF}) coincides with \eq (15) from \cite{iv-NN}.

  Let us consider the master equation (\ref{iv-6FF}). We put
\beq
		E(u) \neq 0 \quad {\rm for}  \quad u \in \left(u_{-}, u_{+}\right), \label{iv-7E} 
\eeq
  where $\left(u_{-}, u_{+}\right)$ is the interval from (\ref{iv-7PN}). 
  Denoting $y = \dot{f}$, we rewrite \eq (\ref{iv-6FF}) as
\bear
		\dot{y} + a(u)y + b(u) = 0, \label{iv-7.Y} 
\ear
  where
\bear
		a(u) = \frac{F(u)}{E(u)}, \qquad   
		b(u) = \frac{G(u)}{E(u)} . \label{iv-7.AB} 
\ear
  The solution to the differential equation (\ref{iv-7.Y}) can be readily obtained 
  by standard methods:
\bearr
		\dot{f} = y = C_0 y_0(u)
\nnn \cm  
	 -   y_0(u)\int_{u_0}^u dw b (w)\left(y_0(w)\right)^{-1},     \label{iv-8.Y} 
\ear
  where $u \in \left(u_{-}, u_{+}\right)$, $C_0$ is a constant, and
\bearr
		y_0(u) = \exp \bigg( - \int_{u_0}^u d v a (v)\bigg)           \label{iv-8.Y0} 
\ear
  is the solution to the homogeheous equation: $\dot{y_0} + a(u)y_0 = 0$.
  Integrating (\ref{iv-8.Y}), we obtain
\bearr
		f =  C_1 + C_0 \int \limits_{u_0}^u dv y_0 (v)
\nnn \qq 
		 - \int_{u_0}^u d v y_0 (v)\int_{u_0}^v dw\,b(w)\left(y_0(w)\right)^{-1}, \label{iv-8.F} 
\ear
  where $C_1$ is a constant. We note that the GAC (\ref{iv-6Phi}) impose restrictions 
  only on $C_0$ and $u_0$ since the function $\Phi(u)$ depends on $\dot{f}$ 
  and $\ddot{f}$. Here $C_1$ is an arbitrary constant.

% ================================
\section{The Schwarzschild  metric test }
% ================================

  Here we test the reconstruction procedure by using the Schwarzschild  metric.

% ----------------------------
\subsection{Basic relations}
% ----------------------------

  Let us start with the simplest case of the Schwarzschild solution with
\bear
	A(u) = 1 - \frac{2\mu}{u} ,  \label{iv-8AC} 
\ear
  where $\mu > 0$ and $u > 2\mu$. In this case, for the master equation (\ref{iv-6FF}) we get 
  for the functions $E(u)$, $F(u)$ and $G(u)$ defined in (\ref{iv-6E}), (\ref{iv-6F6}) and (\ref{iv-6G6}),  
  respectively:
\bearr  
 	E = -\frac{48\mu}{u^2}\left(u - 2\mu\right), \label{iv-8E} 
\yyy 
	 F = \frac{64\mu}{u^3}\left(u - 3\mu\right), \label{iv-8F} 
\yyy	 
	 G = 0. \label{iv-8G} 
\ear
  Solving the master equation $E\ddot{f} + F\dot{f} + G = 0$, we obtain
\bearr
		f(u) = c_1 + c_0 \frac{3}{7}\left(u - 2\mu\right)^{1/3}
\nnn \inch \times		
		\left(u^2 + 3\mu u + 18\mu^2\right) , \label{iv-8fc} 
\ear
  and
\bearr
		\dot{f} = c_0 u^2\left(u - 2\mu\right)^{-2/3}, 
\nnn		 
	\ddot{f} = c_0\frac{4u\left(u - 3\mu\right)}{3\left(u - 2\mu\right)^{5/3}},   \label{iv-8fdot} 
\ear  
  where $c_0$ and $c_1$ are constants, and $u > 2\mu$. Here the integration constants 
  in the solution (\ref{iv-8.F}) are related to those in the solution (\ref{iv-8fc}) as follows:
  $c_0 = C_0(u_0)^{-2} (u_0 - 2\mu)^{2/3}$, $c_1 = C_1$.

  The GAC relation (\ref{iv-6Phi})  in this case reads
\bearr
		 \Phi = \dot{\varphi}^2 = 8 u^{-2} \ddot{f}\left(A - 1\right) 
\nnn \cm 
	= - c_0 u^{-2} \frac{64 \mu (u - 3\mu)}{3(u - 2\mu)^{5/3}} > 0.   \label{iv-8Phi} 
\ear
   It is satisfied if
\beq  
		 c_0 < 0, \quad  {\rm for} \quad u > 3\mu,  \label{iv-8out} 
\eeq  
  and
\beq  
		c_0 > 0, \quad  {\rm for} \quad 2\mu < u < 3\mu.  \label{iv-8in} 
\eeq  
  This means that for $c_0 < 0$ we have a real scalar function at $u > 3\mu$, i.e., out 
  of the photon sphere, obeying
\bear
	\frac{d\varphi}{du} =  8 \eps \Bigl(- \frac{c_0\mu}{3}\Bigr)^{1/2}
	\frac{\left(u - 3\mu\right)^{1/2}}{u\left(u - 2\mu\right)^{5/6}},  \label{iv-8Phi_dout} 
\ear
  $\eps = \pm 1$, which becomes a nonreal complex one for $2\mu < u < 3\mu$, 
  i.e., between the photon sphere and the horizon.

  On the contrary, for $c_0 > 0$ we have a real scalar function at $2\mu < u < 3\mu$, 
  i.e., inside the photonic sphere and out of the horizon, obeying
\bear
	\frac{d\varphi}{du} = 8 \eps \Bigl(\frac{c_0\mu}{3}\Bigr)^{1/2}
	\frac{(3\mu - u)^{1/2}}{u(u - 2\mu)^{5/6}},  \label{iv-8Phi_din} 
\ear
  $\eps = \pm 1$, which becomes a (nonreal) complex one at $u > 3\mu$, i.e. out of 
  the photon sphere. Recall that the radius of the photon sphere in the Schwarzschild 
  solution in the present notations is $3\mu$. In a domain where a ghost is absent, we have 
  a monotonic function $\varphi(u)$, either increasing or decreasing one.

  For  $U(u)$ we obtain the relation (\ref{iv-6UU}) with the following functions
   (\ref{iv-6EU}), (\ref{iv-6FU}), (\ref{iv-6GU}):
\bearr
		 E_U = 8 u^{-2}\left(-\mu\right)\left(u - 2\mu\right), 
\nnnv   
	 F_U = 16u^{-3}\mu\left(u - 3\mu\right), \quad  G_U = 0.   \label{iv-8EFGU}
\ear
  Hence we get the following expression for the potential function:
\bear \nhq
		U(u) =  c_0 \frac{16}{3}  u^{-3}\mu(u - 3\mu) (u - 2\mu)^{-2/3}.  \label{iv-8U} 
\ear 
  According to \eqs (\ref{iv-8U}), (\ref{iv-8out}), and (\ref{iv-8in}), for a given $c_0$ we get: 
  $U(u) < 0$ in a domain where there are no ghosts, and  $U(u) > 0$ in a domain 
  where there is a ghost. The same is true for $\ddot{f}$, see (\ref{iv-8fdot}).

% ------------------------------
\subsection{The scalar field}
% ------------------------------

  Here we consider the scalar field $\varphi = \varphi (u)$ in detail.
  We start with \eqs (\ref{iv-8Phi_dout}), (\ref{iv-8Phi_din}), written in the following form:   
\bearr
	\frac{d\varphi}{du} = \eps b_0
	 \frac{\left(u - 3\mu\right)^{1/2}}{u\left(u - 2\mu\right)^{5/6}},
	  \ \  {\rm for} \ \  c_0 < 0, \  \label{iv-9_1} 
\yyy
	\frac{d\varphi}{du} = \eps b_0 
	\frac{\left(3\mu - u\right)^{1/2}}{u\left(u - 2\mu\right)^{5/6}},
 		\ \  {\rm for} \ \  c_0 > 0,  \ \label{iv-9_2} 
\ear
  where  $\eps = \pm 1$ and
\beq  
		b_0 \equiv 8\left(\frac{\left|c_0\right|\mu}{3}\right)^{1/2}.   \label{iv-9_3} 
\eeq  

  Consider the first case $c_0 < 0$, $ u > 3\mu$. We obtain
\beq  
	 \frac{d \varphi}{du} \sim  \frac{1}{3} \eps b_0 \left(u - 3\mu\right)^{1/2} \mu^{-11/6},   
	 			\label{iv-9_4} 
\eeq  
  as $u \to 3\mu$, and hence 
\bearr  
	\varphi (u) - \varphi (3\mu +0) 
\nnn \cm	
	\sim \eps b_0 \frac{2}{9}\left(u - 3\mu \right)^{3/2} \mu^{-11/6},      \label{iv-9_5} 
\ear  
   as $u \to 3\mu$. For $u \to + \infty$ we obtain another asymptotic relation
\beq  
		\frac{d\varphi}{du} \sim \eps b_0 u^{-4/3},  \label{iv-9_6} 
\eeq  
   which implies
\bearr
    \varphi(+\infty) - \varphi(u)  = \int_u^{+ \infty} d\bar{u}\frac{d\varphi}{d\bar{u}} 
\nnn \cm
    \sim \int_u^{+\infty} \eps b_0 {\bar{u}}^{-4/3}\,d\bar{u}  =  3\eps b_0u^{-1/3},   \label{iv-9_7} 
\ear
  as $u \to +\infty$. We also obtain 
\bearr
	 \varphi(+\infty) - \varphi\left(3\mu + 0\right) 
	 =  \int_{3\mu}^{+ \infty} d\bar{u}\frac{d\varphi}{d\bar{u}}
\nnn 
  = \eps b_0  \int_{3\mu}^{+ \infty} d\bar{u}\frac{\left(\bar{u} - 3\mu\right)^{1/2}}
  	{\bar{u}\left(\bar{u} - 2\mu\right)^{5/6}} = \eps b_0\mu^{-1/3}I_1,        \label{iv-1_9} 
\ear
  where
\beq  
	I_1 = \int_{3}^{+ \infty} dx \frac{\sqrt{x - 3}}{x \left(x - 2\right)^{5/6}}. \label{iv-9_10} 
\eeq  
	By using Wolphram Alpha we find 
\bearr
	I_1 =  \frac{\sqrt{\pi}}{4\Gamma\Bigl(\tfrac{5}{6}\Bigr)}
		\bigg[7\ _2F_1\left(-\tfrac{1}{2}, 1; \tfrac{2}{3}; \tfrac{1}{3}\right)
\nnn \ \ 	\cm	
	 - 18\  _2F_1\left(\tfrac{1}{2}, 1; \tfrac{2}{3}; \tfrac{1}{3}\right) + 18 \bigg]
			 \Gamma\Bigl(\tfrac{1}{3}\Bigr) 
 \nnn    \inch
		  + \pi  \sqrt[6]2 \approx 2.01431.            \label{iv-9_11} 
\ear
  Here and below $_2F_1(x, a; b; c )$ is the hypergeometric function, and 
  $\Gamma(x)$ is the Gamma function.

   Now we consider the second case $c_0 > 0$, $2\mu < u < 3\mu$. We get
\beq  
	\frac{d\varphi}{du} \sim \eps b_0 \frac{\sqrt{3\mu - u}}{3 \mu^{11/6}}, \label{iv-9_12} 
\eeq  
   as  $u \to 3\mu$. This relation implies 
\bearr
		 \varphi\left(3\mu - 0\right) - \varphi(u) =
 				\int_u^{3\mu} \frac{d\varphi}{d\bar{u}} d \bar{u}
 \nnn  \cm
    \sim \frac{2}{9}\eps b_0  \left(3\mu - u \right)^{3/2} \mu^{-11/6},     \label{iv-9_13} 
\ear
   as $u \to 3\mu$.  For $u \to 2\mu$ we get another asymptotic relation,
\beq  
		\frac{d\varphi}{du} \sim \eps b_0 \frac{\mu^{1/2}
		\left(u - 2\mu\right)^{-5/6}}{2\mu},  \label{iv-9_15} 
\eeq  
  which implies 
\bearr%\begin{gathered}
	 \varphi(u) - \varphi\left(2\mu + 0 \right)
		   = \int_{2\mu}^{u}{d\bar{u}} \frac{d\varphi}{d\bar{u}} 
\nnn \cm
		   \sim 3\eps b_0 \mu^{-1/2}\left(u - 2\mu\right)^{1/6},   \label{iv-9_16} 
\ear
  as $u \to 2\mu$.

   We also find another relation,
\bearr
	 \varphi\left(3\mu - 0 \right) - \varphi\left(2\mu + 0 \right)
   				= \int_{2\mu}^{3\mu}{d\bar{u}} \frac{d\varphi}{d\bar{u}} 
\nnn  
  = \eps b_0 \int_{2\mu}^{3\mu}{d\bar{u}} \frac{\left(3\mu - \bar{u}\right)^{1/2}}
  	{\bar{u}\left(\bar{u} - 2\mu\right)^{5/6}} 
		  = \eps b_0 \mu^{-1/3} I_2,   \label{iv-9_17} 
\ear
  where 
\beq  
		I_2 = \int_{2}^{3} dx \frac{\sqrt{3 - x}}{x\left(x - 2\right)^{5/6}}.  \label{iv-9_18} 
\eeq  
  The use of Wolphram Alpha gives us 
\bearr
		I_2 =  \frac{3\sqrt{\pi}}{\Gamma\Bigl(\frac{5}{3}\Bigr)}
		\Bigl[  2\  _2F_1\left(-\tfrac{5}{6}, 1; \tfrac{2}{3}; -\tfrac{1}{2}\right) 
\nnn	\cm	
		-  3\  _2F_1\left(\tfrac{1}{6}, 1; \tfrac{2}{3}; -\tfrac{1}{2}\right)  \Bigr]
		\Gamma\Bigl(\tfrac{7}{6}\Bigr) 
\nnn \inch  
			 \approx 2.61887.      \quad \label{iv-9_19} 
\ear
  In what follows we put for simplicity  
\beq  
 	 \varphi \left(3\mu - 0 \right) = \varphi \left(3\mu + 0 \right) =0, \quad \eps = +1.
     \label{iv-9_20} 
 \eeq  
  Then, for  $c_0 < 0$, the function $\varphi(u)$ is defined on the interval 
  $\left(3\mu, +\infty\right)$. It is  monotonically increasing from $0$ to 
\beq  
     \varphi_1 \equiv \varphi(+\infty) = b_0 \mu^{-1/3} I_1.    \label{iv-9_21_1} 
\eeq  
   For  $c_0 > 0$ the function $\varphi(u)$ is defined on the interval
  $\left(2\mu, 3\mu\right)$. It is  monotonically increasing from $ (- \varphi_2)$ to $0$, where
\beq  
      \varphi_2 \equiv - \varphi\left(2\mu + 0 \right) =  b_0 \mu^{-1/3} I_2.    \label{iv-9_21_2} 
\eeq  

% ---------------------------------------
\subsection{The coupling function}
% ---------------------------------------

  Now we explore the coupling function, assuming the relations (\ref{iv-9_20}). 
  We start with \eq (\ref{iv-8fc}) ,
\bear  \nhq
	 f(u) =  c_0 \frac{3}{7}  \left(u - 2\mu\right)^{1/3}
	 \left(u^2 + 3\mu u + 18\mu^2\right),                \label{iv-10_1}
\ear 
  where we put (without loss of generality)  $c_1 = 0$. Indeed, the inclusion of 
  $c_1 \neq 0$ into the relation (\ref{iv-8fc}) will not contribute to the equations of motion 
  since the Gauss-Bonnet term gives us a topological invariant. We obtain 
\bearr
		f(u) \sim c_0\frac{3}{7}\left(u - 2\mu\right)^{1/3}  28\mu^2
\nnn \inch		
		 = 12 c_0\left(u - 2\mu\right)^{1/3}, \,  \label{iv-10_2} 
\ear  
   as $u \to 2\mu$, 
\beq  
	f\left(3\mu\right) = \frac{108}{7} c_0 \mu^{7/3},  \label{iv-10_3} 
\eeq  
   and 
\beq  
		f\left(u \right) \sim  \frac{3}{7} c_0 u^{7/3} ,   \label{iv-10_4} 
\eeq  
   as $u \to +\infty$. 

  The functions $f(u)$, corresponing to $c_0 > 0$ and $c_0 < 0$,
  are depicted at Fig.\,1  (for  $\mu = 1$ and $b_0 = 1$ ).
% -------------------------------------- fig 1
\begin{figure}[h]
\center{\includegraphics[scale=0.5]{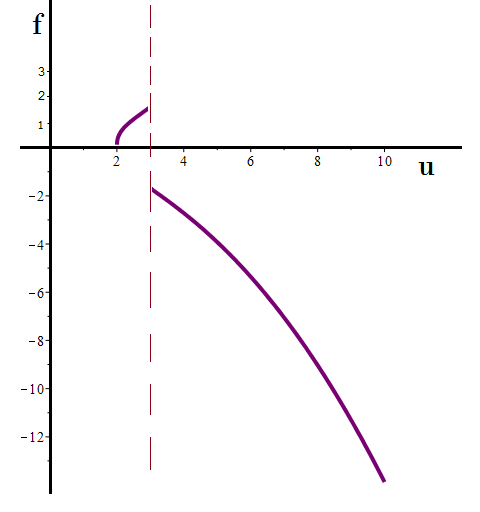}}
\caption{\small
	Two functions $f(u) \equiv f(\varphi(u))$  for  $\mu = b_0 = 1$. 
	Here the $\varphi(u)$ correspondence obeys in our notations: 
	$\varphi(2) = - \varphi_2 < 0$, $\varphi(3) =0$, and  $\varphi(+ \infty) = \varphi_1 > 0$.  } 
\label{iv-Fig_1}
\end{figure}
% --------------------------------------------------

  Let us consider the first case $c_0 < 0$. Due to  
\beq  
 		\varphi_1 - \varphi(u) \sim \const \cdot u^{-1/3}
\eeq  
  as $u \to +\infty$  (see (\ref{iv-9_7}), (\ref{iv-9_20}), and  (\ref{iv-9_21_1})),
  and     (\ref{iv-10_4}), we obtain 
\beq  
 		f(\varphi) \sim - C_{f,1} \left(\varphi_1 - \varphi\right)^{-7}     \label{iv-10_5} 
\eeq  
  as $\varphi \to \varphi_1$. Here $C_{f,1} > 0$ is constant.  

  Now we use the asymptotical relation 
\beq  
		f(u) - f\left(3\mu + 0\right) \sim \dot{f}\left(3\mu + 0\right)\left(u - 3\mu\right),
				\label{iv-10_6}
\eeq   
   as $u \to 3\mu + 0$. We denote
\bearr   
		f_0 = -  f\left(3\mu + 0\right) =  f\left(3\mu - 0\right) 
\nnn \inch		
		=  |c_0|\frac{108}{7}\mu^{7/3} > 0.            \label{iv-10_7}
\ear  
  Due to $\varphi \sim \const \cdot \left(u - 3\mu \right)^{3/2}$ as $u \to 3\mu$ 
  (see (\ref{iv-9_5}), (\ref{iv-9_20})), (\ref{iv-10_6}), and  (\ref{iv-10_7}), we get
\beq  
		f(\varphi) + f_0 \sim - C_{f,+}  \varphi^{2/3},           \label{iv-10_9}
 \eeq  
   as $\varphi \to + 0$.  Here $C_{f,+} > 0$ is a constant 
   proportional to ($\dot{f} \left(3\mu + 0\right) < 0$).  

  Let us consider the second case  $c_0 > 0$. By using the asymptotical relations
\beq  
	f(u) - f\left(3\mu - 0\right) \sim \dot{f} \left(3\mu - 0\right)\left(u - 3\mu\right),
			\label{iv-10_10}
\eeq   
  as $u \to 3\mu $,  and  $(- \varphi) \sim \const  \left(3\mu - u \right)^{3/2}$, 
  as  $u \to 3\mu$, (see (\ref{iv-9_13}), (\ref{iv-9_20})) and (\ref{iv-10_7}), we are led to the 
  following asymptotical relation:
\beq  
 		f(\varphi) - f_0 \sim - C_{f,-}  \left(- \varphi\right)^{2/3},    \label{iv-10_12}
\eeq  
   as $ \varphi \to - 0$. Here  $C_{f,-} > 0$  is a constant, proportional to 
   $\dot{f} \left(3\mu - 0\right) > 0 $.    

  Now we rewrite the asymptotic relation  (\ref{iv-10_2}).
  By using $\varphi + \varphi_2  \sim \const \cdot (u - 2\mu)^{1/6}$   
  (see (\ref{iv-9_16}) and (\ref{iv-9_21_2})), we obtain 
\beq  
		 f(\varphi) \sim C_{f,2} \left(\varphi + \varphi_2\right)^2,       \label{iv-10_13A}
\eeq  
  as $\varphi \to - \varphi_2$. Here  $C_{f,2} > 0$ is constant. 

  For $c_0 < 0$ the coupling function $f(\varphi)$ is defined on the interval $(0, \varphi_1)$.
  It is negative-definite, $f(\varphi) < 0$, and  unbounded since 
  $f(\varphi) \to - \infty$ as $\varphi \to  \varphi_1$.   
  For $c_0 > 0$ the function $f(\varphi)$ is defined on the interval $(- \varphi_2,0)$.
  It is  posive-definite and bounded since  $0 < f(\varphi) < f_0$. At $\varphi \to - \varphi_2$
   it vanishes: $f(\varphi) \to +0$.

% ----------------------------------------
\subsection{The potential function}
% ----------------------------------------

  Now we consider the potential function. Here we keep our agreement 
  (\ref{iv-9_20}). We start with the relation (\ref{iv-8U}) for $U(u)$.  

  At $c_0 < 0$ and  $u > 3 \mu$ we obtain
\beq  
		U(u) \sim  - \left|c_0\right|\frac{16}{3}  u^{-8/3}  \label{iv-11_1}
\eeq  
  as $u \to +\infty$, and
\beq  
		U(u) \sim - \left|c_0\right| \frac{16}{81}  \mu^{-8/3}\left(u - 3\mu\right),     \label{iv-11_2}
\eeq  
  as  $u \to 3\mu$. For  $c_0 > 0$ and $2\mu < u < 3\mu$, we get
\beq      
  		U(u) \sim  - c_0 \frac{16}{81} \mu^{-8/3}\left( 3\mu - u  \right),    \label{iv-11_3}
\eeq  
  as $u \to 3\mu$ and 
\beq   
		U(u) \sim  - c_0 \frac{2}{3}  \mu^{- 1}\left(u - 2\mu\right)^{- 2/3}, \label{iv-11_4}
\eeq  
   as $u \to 2\mu$.

  The functions $U(u)$ corresponing to $c_0 > 0$ and $c_0 < 0$ are depicted in Fig.\,2 
  (for  $\mu = 1$ and $b_0 = 1$).
% ----------------------------------------------- fig 2
\begin{figure}[h]
\centering
\includegraphics[scale=0.5]{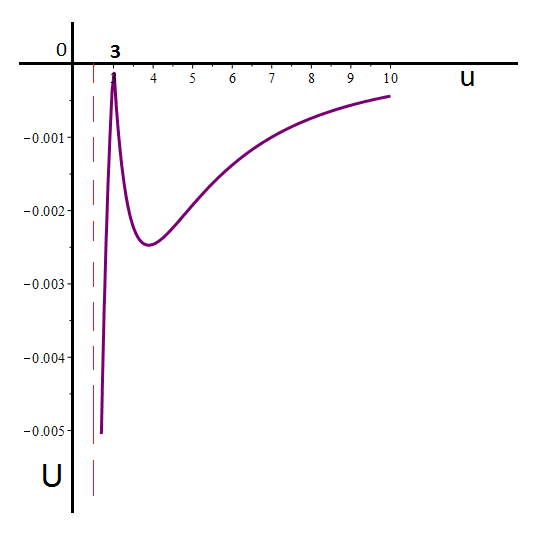}
\caption{\small
		Two functions $U(u) \equiv U(\varphi(u))$ for  $\mu = b_0 = 1$.
		 The function  $\varphi(u)$ obeys: $\varphi(2) = - \varphi_2 $, 
		 $\varphi(3) =0$, and  $\varphi(+ \infty) = \varphi_1$. }
\label{iv-Fig_2}
\end{figure}
% ----------------------------------------------

  In the case $c_0 < 0$, $\mu = 1$ the   point of minimum  is reached at
\beq  
		u_{\ast} = \frac{3\sqrt{33} + 45}{16} \approx 3.8896,     \label{iv-11_5}
\eeq  
   with $U\left(u_{\ast}\right) \equiv U_{\ast} = -(16/3)|c_0| N_{\ast} $, where\\
   $N_{\ast} \approx  0.0098908$, obtained as
\bearr
	N_{\ast} = N_1/N_2,
\nnn	
	N_1 = 	
	\Bigl(11072\sqrt{33} + 49344\Bigr)\Bigl(\frac{3\sqrt{33} + 45}{16} - 2\Bigr)^{1/3},
\nnn	
	N_2 = \frac{\left(3\sqrt{33} + 45\right)\left(645165\sqrt{33} + 3765123\right)}{16} 
\nnn \cm	
	- 1290330\sqrt{33} - 7530246.                                  \label{iv-11_6}
\ear

  Now we consider the  potential function in terms of the original variable, i.e.,  $U(\varphi)$.
  For $c_0 < 0$, we find
\beq  
		U(\varphi) \sim C_{U,1} (\varphi_1 - \varphi)^8,         \label{iv-11_7}
\eeq   
  as $\varphi \to \varphi_1$, where $C_{U,1} < 0 $  is constant, and
\beq  
			U(\varphi) \sim C_{U,0} \varphi^{2/3},   \label{iv-11_8}
\eeq  
   as $\varphi \to + 0$, where  $C_{U,0} < 0 $ is constant. For $c_0 > 0$ we obtain 
\beq  
		 U(\varphi) \sim C_{U,0} \left(-\varphi \right)^{2/3},  \label{iv-11_9}
\eeq  
   as $\varphi \to - 0$,  and
\beq  
		 U(\varphi) \sim C_{U,2}  (\varphi + \varphi_2)^{-4}, \label{iv-11_10}
\eeq  
  as $\varphi \to -\varphi_2$, where  $C_{U,2} < 0$ is constant. 

  We see that in both cases  $U(\varphi) < 0$. For $c_0 < 0$ we obtain 
\beq  
	U(\varphi) \geq U\left(\varphi_{\ast}\right) = U_{\ast},   \label{iv-11_11}
\eeq  
  where $\varphi_{\ast} = \varphi\left(u_{\ast}\right) \approx 0,01145$, 
  i.e., the potential $U(\varphi)$ is bounded. For $c_0 > 0$ we get 
   $U(\varphi) \to - \infty$
  as $\varphi \to -\varphi_2$, i.e., potential is unbounded.

% =======================
\section{Conclusions}
% =======================

  We have studied the 4D gravitational model with a real scalar field $\varphi$, 
  Einstein and  Gauss-Bonnet terms. The action contains the potential term $U(\varphi)$ 
  and the Gauss-Bonnet coupling function $f(\varphi)$. For a special (static) spherically 
  symmetric metric $ds^2 = du^2/A(u)  - A(u)dt^2 + u^2 d\Omega^2$,
  with a given redshift function $A(u) > 0$  ($u > 0$ is a radial coordinate), we have verified  
  the so-called reconstruction procedure suggested by Nojiri and  Nashed \cite{iv-NN},
  according to which there exists a pair of $U(\varphi)$ and  $f(\varphi)$, described by certain
  implicit relations, which leads us to exact solutions to the equations of motion with  a given 
  metric governed by  $A(u)$. Here we have confirmed that all relations in Ref.\,\cite{iv-NN} 
  for $f(\varphi(u))$ and $\varphi(u)$ are correct, but the expression for $U(\varphi(u))$ 
  contains a typo which is eliminated in this paper.

  We have applied the reconstruction procedure to the external Schwarzschild black hole 
  metric with the gravitational radius $2 \mu > 0$ and $u > 2 \mu$. Using  the ``no-ghost''
  restriction (i.e.,  reality of $\varphi(u)$), we have found two sets of $(U(\varphi), f(\varphi))$. 
  The first one gives us the Schwarzschild metric defined at $u > 3 \mu$, and the second 
  one describes the Schwarzschild metric defined for $ 2 \mu < u < 3 \mu$.
  In both cases the potential $U(\varphi)$ is negative. For the first set  $(U(\varphi), f(\varphi))$
  with $\varphi \in (0, \varphi_1)$, the potential $U(\varphi)$ is bounded, and the coupling 
  function $f(\varphi) < 0$ is unbounded, while for the second  set  $(U(\varphi), f(\varphi))$
  with $\varphi \in (-\varphi_2, 0)$ the potential $U(\varphi)$ is unbounded, and the coupling 
  function $f(\varphi) > 0$ is bounded.

  It should be noted that here $3 \mu$ is the radius of the photon sphere, which means 
  that the two domains, where we have real scalar field solutions, are separated by the 
  photon sphere. The general analysis of Ref.\,\cite{iv-NN} and its application to the Hayword
  black hole solution indicates the possibility to solve the ghost avoidance problem at least 
  locally, i.e,. in two ranges of the radial variable: $(r_h, r_{1,*})$ and $(r_{2,*}, + \infty)$, 
  where $r_h$ is the horizon radius, and $r_h < r_{1,*} < r_{2,*}$. The problem of enlarging 
  these intervals such that $r_{1,*} = r_{2,*} = r_{*}$ was not studied in Ref. \cite{iv-NN}. 
  This problem may be addressed in the forthcoming publications devoted to the reconstruction 
  procedure for a general class of static spherically symmetric metrics 
  $ds^2 = du^2/A(u) - A(u)dt^2 + C(u) d\Omega^2$
  (with the areal function $C(u) > 0$), with application to dilatonic black holes, e.g., 
  those from \cite{iv-MBI,iv-IMNT}.
  
  We note also that  the  reconstruction problem  for general sperically symmetric metrics which appear in sEGB model was explored (up to  resolving of the ghost avoiding problem)
  in Ref.    \cite{iv-NojNash}. Meanwhile, it was shown  in Ref.  \cite{iv-BrBI} that arbitrary static spherically symmetric metric may be presented (though, in local parts) as a solution to equations of motion of some scalar tensor theory belonging to the class of Bergmann et al. In Ref.  \cite{iv-Bronnikov}  and in  some other papers the authors were able to present an  arbitrary static spherically symmetric metric obeying  $R^0_0= R^1_1 $ as  coming from  a ``magnetic'' solution of certain $GR + NED$ theory ($NED$ means nonlinear electrodynamics).

\Funding
{The research was funded by RUDN University, scientific project number FSSF-2023-0003.    }

\ConflictThey

% ===========================================

 \end{document}